\documentstyle[epsfig]{article}
 \def\thebibliog{\noindent{\bf References}
 \list{[\arabic{enumi}]}{\settowidth\labelwidth{99}\leftmargin\labelwidth
\advance\leftmargin\labelsep\itemsep\parskip
\usecounter{enumi}}\def\newblock{\hskip .11em plus .33em minus .07em}
 \sloppy\clubpenalty4000\widowpenalty4000\sfcode`\.=1000\relax}

\def\Asym{\mbox{{$A$\raisebox{-0.25em}{{\small{\small{\sl 0}}{\sl n}}}}}}

\def\dsdt{{\rmd \sigma\!/\!\rmd {\rm t}}}

\newcommand{\piNpiN}{\pi N \rightarrow \pi N}

\newcommand{\siml}{\parbox{0.8em}{\raisebox{0.3ex}
	{$<$}\hspace{-0.8em}\raisebox{-0.3em}{$\sim$}}}

\def\degree{\mbox{$^\circ$}}

\def\rmd{{\rm d}}
\def\etal{{\it et al.}}

\def\plab{\mbox{{$p$\raisebox{-0.25em}{{\small{\sl lab}}}}}}
\newcommand{\pbar}{\overline{p}}
\newcommand{\pbarp}{\overline{p}p}
\newcommand{\pbarppiopio}{\overline{p} p  \rightarrow \pi^0 \pi^0 }
\newcommand{\pbarppipi}{\overline{p} p  \rightarrow \pi  \pi }

\begin{document} 
\begin{center}

\title
{Exclusive Hadronic Reactions at High $Q^2$\footnote{
Talk at the TJNAF workshop on the Transition from Low to High $Q$ 
Form Factors, Athens, GA, Sept 17, 1999}}


\author{
F. Myhrer \\
Department of Physics and Astronomy, 
University of South Carolina, \\ 
Columbia, SC29208, USA  }


\maketitle


{\bf Abstract.} 

\end{center}

\baselineskip=0.7cm

The observed scaling laws for large angle exclusive hadronic 
reactions are successfully accounted for by the short-distance 
QCD quark interchange processes. 
We focus on three hadronic reactions which shows evidence
for a slight oscillatory deviation from the expected scaling behaviour. 
Possible explanations for these  oscillations and 
connections to spin observables will be mentioned.  
Better data from excisting facilities 
(or a possible future KEK proton accelerator, JHP) 
can clarify the theoretical situation.

\section{ Introduction}

In high energy exclusive hadronic reactions at  
high momentum transfers, $Q^2$, 
we explore the short distance $\sim 1/Q$ 
hadronic interactions and quark dynamics 
that are expected to give the dominant 
features of the cross sections. 
The measured cross sections of various exclusive hadronic 
reactions at large $Q^2$ have successfully 
confirmed the expected scaling laws
of short distance QCD and the quark interchange model ($QIM$) 
discussed in, e.g. Ref.\cite{Sivers}. 
The expected power law fall-off of the differential cross sections 
at large angles are predicted to have the following behaviour: 
$\dsdt (\theta \sim 90 \degree )$ 
$\sim s^{-N}$, 
where $N$ depends on the specific reaction. 
For example, the measured $\dsdt$  at $90\degree$ for 
$pp$ and $\pi^- p$ elastic scattering 
have confirmed the power law fall-off where $N$ equals 10 and 8, 
respectively. 
In Fig.1 we show the measured $\pi^- p$ elastic scattering 
where the power law
prediction of $QIM$ is the obvious dominant feature 
for   $\dsdt$ versus ln($s$). Note that $\dsdt$ changes by 
ten orders of magnitude 
in this figure. For the two highest energies in Fig.1, 
\plab = 20 and 30 GeV/c \cite{Oslo}, the cross section at $90\degree$ 
is tiny making the measurement extremely hard, and this is 
reflected in the large error bars at these two energies. 

The data for $pp$ and $\pi^- p$ elastic scattering  
indicate that we have to refine $QIM$ to account for some 
intriguing features of the data. The phenomenon of interest as observed in 
$pp$ elastic scattering is the ``oscillations"  
with energy  of $\dsdt$ at 90\degree\ about the expected 
smooth power law fall-off \cite{Sivers}. 
This puzzling behaviour has been known for some time and is 
seen  in Fig.2. 
In Fig.2 the ``scaled'' cross section, 
$s^{10} \dsdt$ at $90\degree$, as a function of ${\rm ln}(s)$ is shown. 
If $QIM$ were perfect 
we expect a constant ``scaled" cross section (within error bars).  
Instead 
we clearly see indications of  ``oscillations" about a horizontal line.

\medskip

\begin{figure}
\begin{center}
\epsfig{file=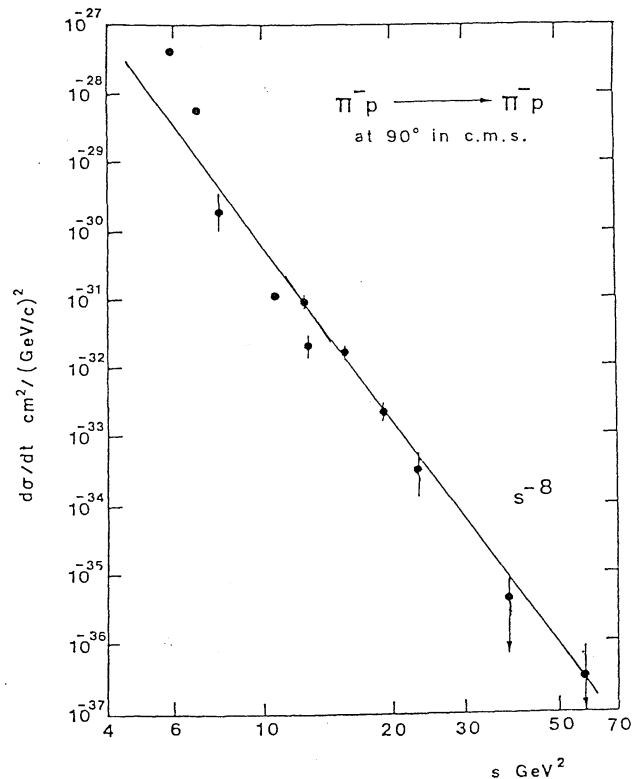,width=9cm}
\vspace*{-1cm}
\end{center}
\caption[]{ The differential cross section $\dsdt$ at $90\degree$ 
for $\pi^- p$ elastic scattering as a function of $s$. 
Figure from Ref.\cite{Oslo}. }
\end{figure}

\medskip

\begin{figure}
\begin{center}
\epsfig{file=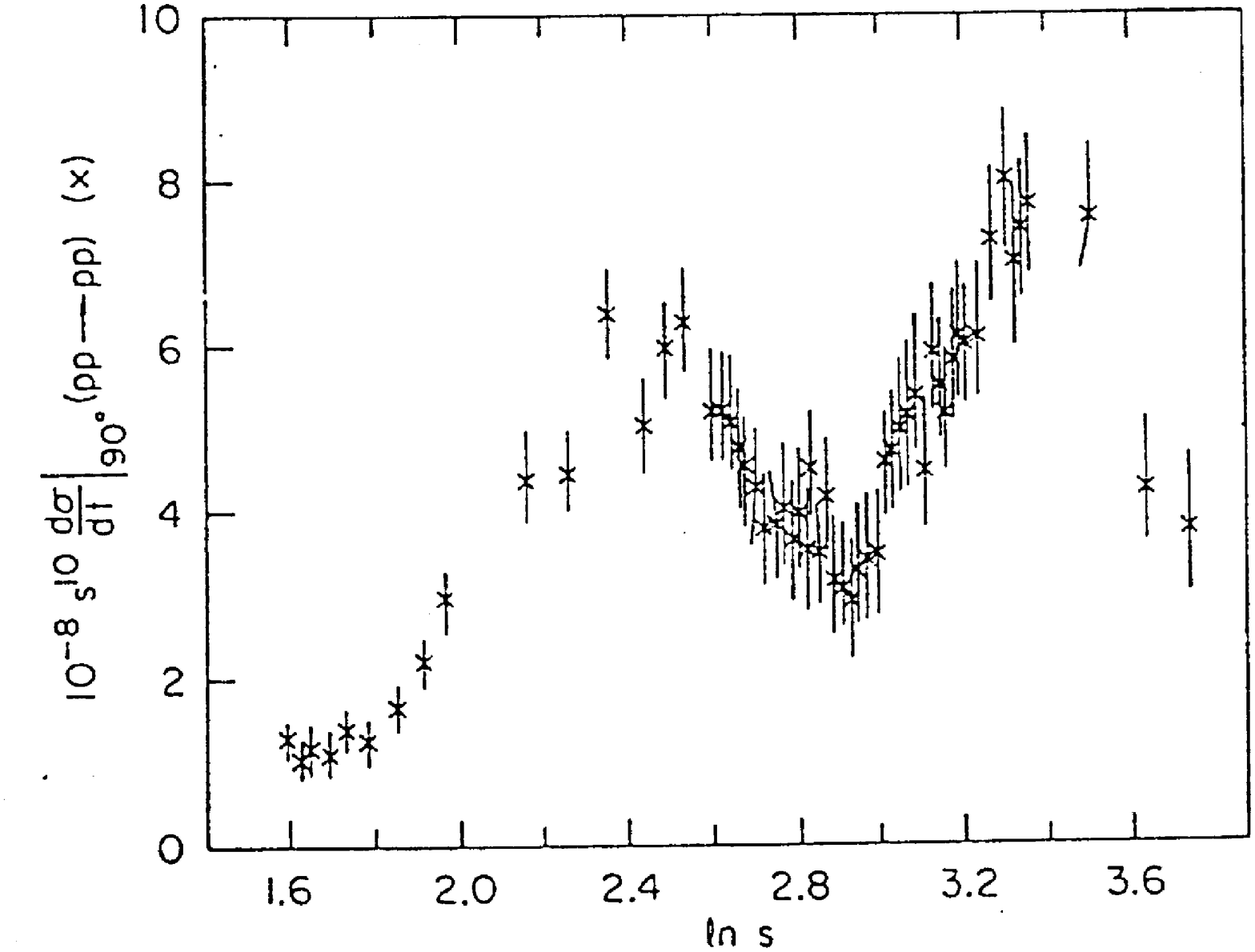,width=7cm}
\end{center}
\caption[]{ The elastic $pp$ cross section 
at 90$\degree$ scaled by $s^{10}$ as a
function of ln($s$/1 GeV$^2$). Figure  from the review of 
Sivers \etal\ \cite{Sivers}. }
\end{figure}

One question which is natural to ask is if these 
``oscillations" are observed in other exclusive hadronic reactions. 
We turn to two other exclusive reactions which 
possibly are even better processes to study, 
namely $\pi^- p$ elastic scattering 
and the reaction $\pbarppipi$.  
These two reactions are in several ways simpler theoretically
than the $pp$ elastic scattering; they could show the 
``oscillation" phenomena more strikingly,  phenomena
seen but inadequately mapped out in the $pp$ case. 
It might even be 
possible to measure both the cross section $\dsdt $ at high 
momentum transfer and the asymmetry  
$\Asym$ over the full kinematic range. 
Keep in mind that $\Asym$ is
large for $\pbarppipi$ at $\plab$ $\simeq$ 2 GeV/c. 
The two reactions $\pi^-p \rightarrow \pi^-p$ and $\pbarppipi$ 
are simpler than $pp$ elastic scattering
if only because they have fewer helicity amplitudes.
A consequence of this is that there is less chance of  averaging
out the ``oscillatory"  effects partly because the 
QCD perturbation theory diagrams will involve fewer quark lines.
An experimental consideration is that  presently the annihilation reaction 
$\pbarppiopio$ has recently been measured and 
further  measurements are possible at 
the antiproton accumulator at Fermilab. 

\section{The ``Oscillations" about the scaling laws} 

The ``oscillations" clearly goes beyond the 
$QIM$ scaling law predictions and
requires some refinements of the arguments leading to the $QIM$ 
scaling laws. 
The ``oscillations" of $pp$ elastic scattering, Fig.2, 
might also be observed in two other exclusive hadronic reactions. 
In Fig.3 the scaled cross section at 90\degree for elastic $\pi^-p$ scattering, 
$s^8 \dsdt$  is shown. 
This figure is taken from Blazey's thesis \cite{Blazey}. 
In Fig.3  the \plab = 30 Gev/c point of Fig.1  is not shown.
This $\plab$ = 30 GeV/c point has too large errorbars to be useful. 
It is however clear from Fig.3 
that we need more accurate data points at the higher energies, 
ln($s$) $>$ 3 ($\plab >$ 10 GeV/c) to draw any firm conclusions. 

\medskip

\begin{figure}
\begin{center}
\epsfig{file=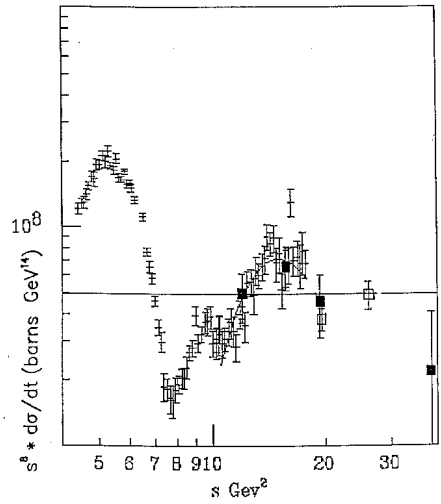,width=6cm}
\end{center}
\caption[]{ The cross section for elastic $\piNpiN$ scattering at 90$\degree$ 
scaled by a factor $s^8$ as a function of $s$.
Figure  from Ref. \cite{Blazey}. }
\end{figure}

At Fermilab the E760 collaboration has measured the differential cross section 
for the reactions $\pbarppiopio$,  $\pbarp \rightarrow \pi^0 \eta$,
$\pbarp \rightarrow \eta \eta$  and 
$\pbarp \rightarrow \pi^0 \gamma$ \cite{E760}. 
The $\dsdt$ at 90\degree for the $\pbarppiopio$ reaction should  show 
a power law fall-off 
$\sim s^{-8}$, similar to $\pi^-p$ elastic scattering.\footnote{
Data for $\pbarp \rightarrow \pi^+ \pi^-$ can be found in 
Refs.\cite{Oslo1,Oslo2,Eisen} }
In Fig.4 the measured $\dsdt$ at 90\degree has been multiplied 
by $s^8$  
and  plotted as a function of ln($s$) \cite{Zioulas}. 
The data points present  evidence for some fluctuations. 
The highest $\pbar$ energy at the Fermilab antiproton accumulator 
corresponds to the ln($s$) $\simeq$ 2.9 point in Fig.4. 
It clearly would be desireable to have a more accurate measurement 
at this energy as well as at other non-measured ln($s$) 
energies of Fig.4 before we can infer anything about a possible 
``oscillation" for this reaction. 

\medskip

\begin{figure}
\begin{center}
\epsfig{file=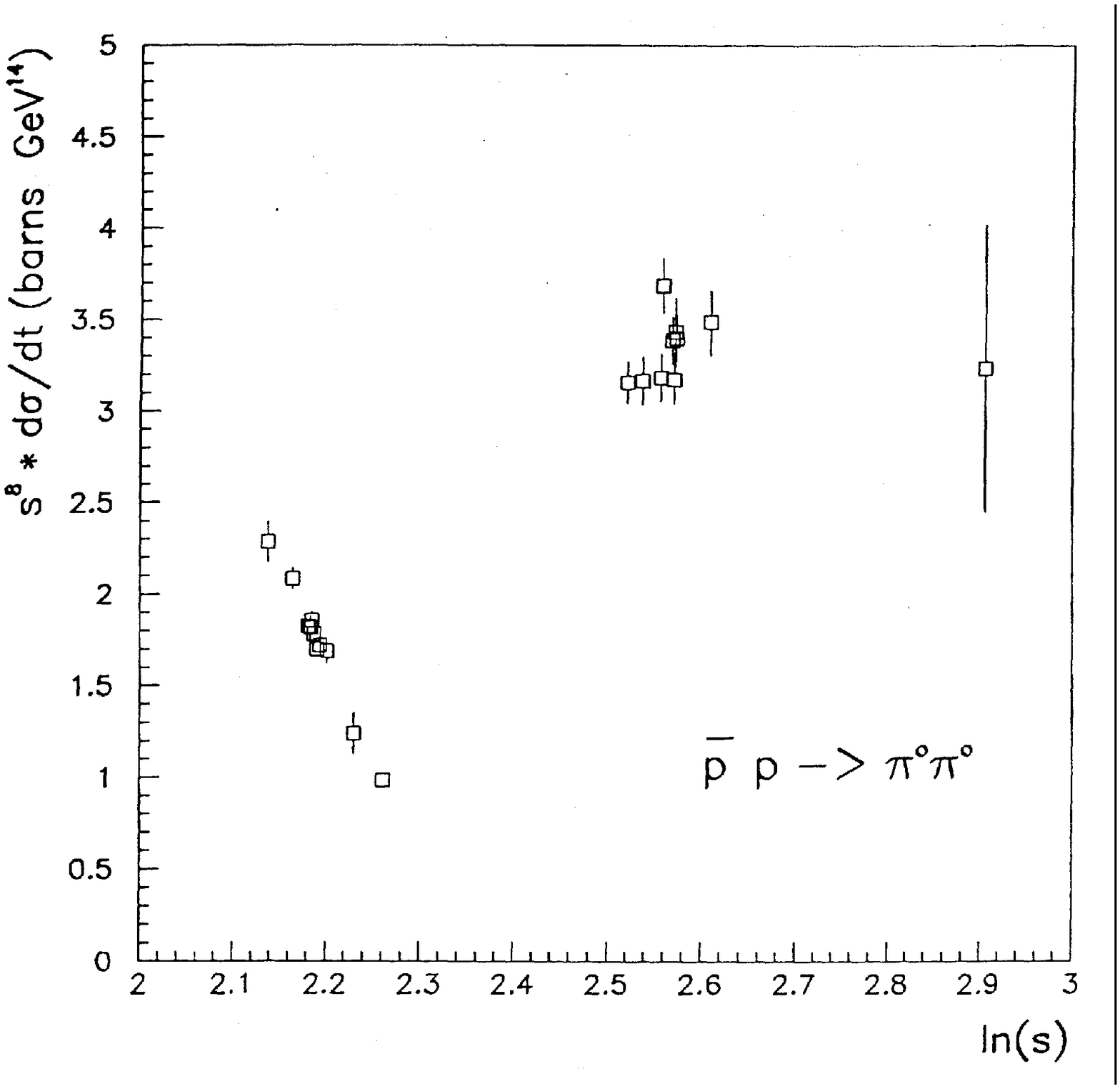,width=6cm}
\end{center}
\caption[]{ The ``scaled" cross section 
s$^8$ $\frac{d\sigma}{dt}(90\degree)$ 
for the reaction $\pbarppiopio$ as a function of ln($s$). 
Figure from Ref.\cite{Zioulas}. }
\end{figure}

The questions related to the above figures are: 
\begin{enumerate}
\item 
Do we observe ``oscillations" in $s^N \dsdt (\theta = 90\degree )$
versus ln($s$) for all three hadronic reactions discussed?
\item
If there are ``oscillations" in all three reactions, 
do the ``oscillations" have
the same ``period" in ln($s$)?
\end{enumerate}
We clearly need more measurements to make progress in the physics understanding
of these phenomena.

\section {Spin Observables}

Another useful observable relevant in the search 
for corrections to the $QIM$ is
the analyzing power, $\Asym$. 
The reason for investigating $\Asym$ is the following. 
The two exclusive reactions $\pi^-p$ elastic scattering and $\pbarppipi$ 
are described by the two helicity amplitudes, $f_{++}$ and $f_{+-}$ whereas 
$pp$ elastic scattering requires five helicity amplitudes 
(only three independent ones at 90\degree ). 
In short distance QCD 
considerations  ($QIM$) helicity is conserved. 
A consequence of helicity conservation is that 
$f_{++}$ = 0 for the reaction $\pbarppipi$. 
For $pp$ elastic scattering helicity conservation 
implies that the two helicity-flip  
amplitudes $\phi_2 (++,--)$ and $\phi_5(++,+-)$ both equal zero. 
This means  that the analyzing power  $A_{0n}$ = 0 
for $pp$ elastic scattering and for $\pbarppipi$. 

We know that  
the  measured analyzing power  

\hspace{5cm} $\Asym  = \frac{d \sigma (\uparrow) - d \sigma (\downarrow)}
{ d \sigma (\uparrow) + d \sigma (\downarrow)} $

\noindent
for $pp$ elastic scattering has 
a significant asymmetry 
even at fairly high momentum transfers \cite{Crabb,Cameron} 
A non-zero $A_{0n}$ in $pp$ elastic scattering implies that 
the $pp$ helicity  amplitude $\phi_5(++,+-)$ must be  non-zero 
and  only the refinements of $QIM$ 
will give a $\phi_5(++,+-)$ $\ne$ 0. 
The same argument applies to 
$A_{0n}$ for the reaction  $\pbarppipi$. Helicity conservation 
implies $f_{++}$ = 0,  and since 
\begin{eqnarray*}
\Asym = 2 \Im m(f_{++}^* f_{+-}) / \left(| f_{++} |^2 +  |f_{+-} |^2 
\right) \; , 
\end{eqnarray*}
a measured non-zero $\Asym$ means we have to augment $QIM$ with 
helicity non-conserving processes. 
The $\pbarppipi$ reaction
has a large asymmetry, $\Asym \simeq $ +1, for 
$p_{lab} \siml$ 2.2 GeV/c (ln($s$) $\siml$ 1.8) \cite{Carter}
and should be measured at higher energies. 
For $\pbarppiopio$ a 
polarized gas jet target may allow measuring $\Asym$ 
even when the cross section is getting small.

Another measured spin observable shows ``structure" 
in $pp$ elastic scattering at high $Q^2$, namely 
the beam target spin correlation $A_{nn}$ at $90\degree$ 
up to s = 26 GeV$^2$
\cite{Crosbie81}.
$A_{nn}$ is not predicted to be zero even if the short
distance processes dominate \cite{Farrar,Lipkin}, but one would
not expect large variations of $A_{nn}$ in a regime 
where one can use perturbative
QCD unless there were interference effects.
For $pp$ elastic scattering at 90\degree\ there is a sum rule 
(independent of QCD) which says \cite{Lipkin}:
\begin{eqnarray*} 
A_{nn} + A_{ss} + A_{ll} = 1 \; , 
\end{eqnarray*}
where $n$ signify a spin polarized normal to the scattering plane, 
and where $s$ is spin polarized perpendicular to the 
proton momentum in the scattering plane and $l$ is 
longitudinal  polarized spin. 
This implies that even at the highest energies there are 
non-zero $pp$ spin observables.  

\section{Theoretical ideas for corrections to  $QIM$}

The observation by Brodsky and de Teramond \cite{Brodsky88} that 
for $pp$ elastic scattering  we can have possible dibaryon resonances 
associated with  the charm threshold, e.g. $pp$ $\rightarrow$ 
$\Lambda^+_c \overline{D}^0 p$ and a single resonance amplitude 
with $J = L = S =1$ will give $A_{nn}$ = +1. Interference with the $QIM$ 
amplitude will then produce a value for $A_{nn}$ $<$ 1.
They tie the ``structure" of $A_{nn}$ to the opening of the 
$c \overline{c} $ threshold and they also show that the resonance 
can  produce ``oscillations" around the smooth power scaling 
fall-off for the differential cross section. 
We do expect resonance phenomena at the opening of new thresholds and
we note that the possible ``oscillations" for the two other reactions
$\pi^- p$ elastic scattering and $\pbarppipi$  appear 
at about the same energies. The task   is to investigate possible
 $\overline{c} c$ threshold resonances for the two other reactions 
 $\pi^- p$ elastic scattering and $\pbarppipi$.

Another process which could interfer with and 
produce some corrections to 
the $QIM$ amplitudes, is the Landshoff process 
\cite{Lands74}. 
It has been proposed that this interference also could be an 
explanation of the $pp$
``oscillations"  \cite{Ral88,Ral86}, but this 
requires some reworking in light of recent
developements \cite{BS89}.
The Landshoff process allows ``independent" pairs of quarks to interact 
via hard gluon exchange and the different interacting quark pairs 
can be separated by a non-negligible  impact parameter.
At high $Q^2$ the longitudinal dimension and one transverse dimension 
will be of the order $1/Q$. The third dimension will be influenced by the size 
(r.m.s. radius) of the hadrons of the reaction. 
Naively speaking the Landshoff process suggest an energy behaviour 
of $\dsdt (\theta = 90\degree )$ 
$\sim s^{-N_L}$ where $N_L$ $< N$ even when QCD  
radiative corrections to the Landshoff process is included 
\cite{BS89,Mueller}. 
These radiative corrections are calculated in perturbative QCD 
or derived heuristically. If $N_L$ $< N$ 
then the Landshoff process should dominate at high energies. 
However, since the transverse dimensions or sizes of the hadrons 
have to be considered,  
meaning soft QCD processes are implied in the calculations,  
we should take 
the calculations of $N_L$ with a grain of salt.

The Landshoff amplitudes contain soft QCD processes
where a propagator is (almost) on-shell
(Sudakov form factors), 
called "the Landshoff pinch", see e.g. Ref.\cite{Mueller}.
The crucial realization is that 
the radiative corrections give the 
quark-quark scattering amplitude  an energy dependent 
phase \cite{Ral88}. 
The Landshoff  process therefore allows for helicity flip 
in the reactions due to the ``soft" transverse 
dimensions of the hadrons 
\cite{PennState,CCM}. 
This energy dependent phase acts at what one might call
medium-high energy, 
although at asymptotic energies the
phase becomes energy independent as stressed by Botts and Sterman
\cite{BS89,Botts90}.

\section{Conclusions}

The power law predictions of $QIM$ are highly successful. 
The challenge is to understand the corrections to the scaling laws. 
If the observed  energy ``oscillations" 
have their origin in ``short" distance quark dynamics, 
this ``oscillatory" behaviour should manifest itself in many exclusive 
hadronic reactions.
The question being asked is if the "scaled" cross sections
for {\it both} $\pi N$ elastic scattering {\it and} the annihilation reaction
$\pbar p \rightarrow$ {\it two pseudoscalar mesons}
do ``oscillate" with energy
similar to what is observed for $pp$ elastic scattering.

A further test of the ideas presented here would be to measure 
$\Asym$ for the two  
reactions $\pi^- p$ elastic scattering and $\pbarppipi$. 
With only two helicity amplitudes one can then expect to
disentangle completely the phases and the energy dependence
of both reactions and  
the energy ``oscillations" should  be evident in the amplitudes.
Experimentally, since the
asymmetry is very large at low energies,
$p_{lab} \approx$ 2 GeV/c 
\cite{Carter}, the annihilation reaction $\pbarppipi$ might be 
the best reaction to measure $\Asym$.
We expect the geometric hadronic impact parameter 
ideas used to explain this large asymmetry for $\plab$ $\siml$ 2 GeV/c
\cite{TMK}  to break down at higher energies
when the short distance  QCD regime of exclusive
hadronic reactions is reached.
The onset of the perturbative QCD regime
may be signaled by 
a significant change in the energy and 
angular variation of the asymmetry. For example,
the very large $\Asym$ at 90$\degree$ at $p_{lab} \approx$ 2 GeV/c 
will become smaller and
might ``oscillate" with increasing energy if
the QCD phenomenology outlined above is reasonable.

As stated above it is necessary to complement the measured points of Fig.4 
\cite{E760} to make certain we observe  
``oscillations" in $s^8 \dsdt$ for the reaction $\pbarppiopio$.  
The E835 collaboration at Fermilab could  
contribute with new measurements of this reaction. 
In addition, the measurement of \Asym\ 
for \plab $>$ 2 GeV/c is expected to be extremely useful not only for
a better understanding of
the nature of the extraordinarily large asymmetry 
of $\pbarppipi$ observed \cite{Carter}, 
but also for monitoring the possible onset of perturbative QCD.

\medskip

This work is supported in part by NSF grant no. PHYS-9602000.

\medskip

\begin{thebibliog}

\bibitem{Sivers} D. Sivers, S.J. Brodsky, and R. Blankenbecler,
Phys. Rep. {\bf 23}, 1 (1976)

\bibitem{Oslo} C. Baglin \etal\ , Nucl. Phys.  {\bf B216}, 1 (1983)

\bibitem{Blazey} G. Blazey, Ph. D. Thesis,  Univ. Minnesota (1987)
unpublished.

\bibitem{E760} The E760 collaboration: T.A.  Armstrong et al. 
Phys. Rev. D {\bf56}, 2509 (1997); 
J. Reid, Ph.D. thesis (1993) Penn State Univ.

 \bibitem{Oslo1} T. Buran et al., Nucl. Phys. ${\bf B116}$, 51 (1976) 

\bibitem{Oslo2} \AA .  Eide \etal , Nucl. Phys. ${\bf B60}$, 173 (1973)

\bibitem{Eisen} E. Eisenhandler et al., Nucl. Phys. ${\bf B96}$, 109 (1975)

\bibitem{Zioulas} G. Zioulas, LEAP94 Conf. Proc., eds. G. Kernel et al. 
 (World Scientific, 1995) p.647;  
 and private communatication. 



\bibitem{Crabb} D.G. Crabb \etal , Phys. Rev. Lett. $\underline{65}$ (1990) 3241
\bibitem{Cameron} P.R. Cameron \etal , Phys. Rev. $\underline{D32}$ (1985)3070

\bibitem{Carter} A.A. Carter et al., 
Nucl. Phys. {\bf B127}, 202 (1977); 


\bibitem{Crosbie81} E.A. Crosbie et al., 
Phys. Rev. D {\bf 23}, 600 (1981) 

\bibitem{Farrar} G. Farrar, S. Gottlieb, D. Sivers and G. Thomas,
Phys. Rev. D {\bf 20}, 202 (1979) 
\bibitem{Lipkin} S.J. Brodsky, C.E. Carlson and H. Lipkin
Phys. Rev. D {\bf 20}, 2278 (1979) 

\bibitem{Brodsky88} S.J. Brodsky and G.F. deTeramond, Phys. Rev. Lett.
$\underline{60}$ (1988) 1924
\bibitem{Lands74} P. V. Landshoff, Phys. Rev. $\underline{D10}$ (1974) 1024;
P. Cvitanovic, $ibid.$ $\underline{10}$ (1974) 338

\bibitem{Ral88} B. Pire and J.P. Ralston, Phys. Lett. 
{\bf B117}, 233 (1982); 
J.P. Ralston and B. Pire, Phys. Rev. Letters {\bf 49}, 1605 (1982)
\bibitem{Ral86} 
J.P. Ralston and B. Pire, Phys. Rev. Letters {\bf 57}, 2330 (1986)
\bibitem{BS89} J. Botts and G. Sterman, Nucl. Phys. $\underline{B325}$ (1989)62

\bibitem{Mueller} A. H. Mueller, Phys. Rep. {\bf 73}, 237 (1981)


\bibitem{PennState} J.P. Ralston and B. Pire, AIP Conf. Proc.
{\bf 223}, 228 (1991) 
\bibitem{CCM} C. E. Carlson, M. Chachkhunashvili and F. Myhrer,
Phys. Rev. 1992, {\bf D46}, 2891.
\bibitem{Botts90} J. Botts, Nucl. Phys. $\underline{B353}$ (1990) 20
\bibitem{TMK} S. Takeuchi, F. Myhrer and K. Kubodera, 
Nucl. Phys. {\bf A556}, 601 (1993)


\end{thebibliog} 


%

\end{document}